# Comment on "Dynamical Reduction Models"
# by A. Bassi and G.C. Ghirardi

**Fred H. Thaheld**

fthaheld@directcon.net

**Abstract**

Bassi and Ghirardi have developed a new theory to address the measurement problem based upon non-linear and stochastic modifications of the Schrödinger equation which has been given the name Quantum Mechanical Spontaneous Localization or QMSL, with one of the emphasis dealing with reduction or wavefunction collapse within the nervous system. Analysis of this portion of their theory reveals that it faces several problems.

In a recent paper Bassi and Ghirardi have reviewed in detail their new approach attempting to overcome the difficulties that standard quantum mechanics meets in accounting for the measurement problem [1]. The measurement problem arises because standard quantum mechanics contains two dynamical evolution principles, one governed by the linear, deterministic Schrödinger equation, and the other by the von Neumann non-linear indeterministic projection postulate or wavefunction collapse. These principles are radically different and contradict each other. Their new approach is based on non-linear and stochastic modifications of the Schrödinger equation, and has been given the name of Quantum Mechanical Spontaneous Localization or QMSL.

They have attempted to apply this QMSL theory to both the inanimate world of particles and the animate world, emphasizing in Sec. 12.3, 'Reduction within the nervous system'. It is felt that this animate part of their theory is incorrect, based upon the following analysis, quoting first from Sec. 12.3.

"The very possibility of considering QMSL as yielding a unified description of all physical phenomena, rests on the fact that one can show that the physical processes occurring in *sentient* beings, leading to definite perceptions, involve a displacement of a sufficient number of particles over appropriate distances to allow the reduction to take place within the perception time". They then proceed to describe the visual perception process and, the three main cascades of events that take place *following* the absorption of one photon in a photoreceptor cell of the retina: (i) the multiplicative chain in the photoreceptor or rod cell, (ii) transmission of the electrical signals along the fibers of the optic nerve and (iii) the excitation of neurons in the cortical visual area. Reduction is then supposed to take place when the number of particles reaches a certain level. I have deliberately left out all the details and numbers used by them regarding these three main cascades of events to not only keep this analysis simple but, because it turns out they are not relevant anyway to the issue before us.

The problem with this theory is that the collapse of the wavefunction has already taken place when the photon is absorbed in a photoreceptor or rod cell of the retina [2]. The photon is actually absorbed each time in a stochastic fashion by only one of the ~ $10^8$ rhodopsin molecules in each rod cell. Each rhodopsin molecule consists of opsin and a retinal chromophore or light harvesting molecule [2]. As has been previously shown, when a photon is absorbed by a retinal molecule, an electron in the highest $\pi$ orbital is immediately excited to a $\pi^*$ orbital, which means that a jump has been made from one orbital to another and, that a collapse of the wavefunction has taken place [2]. One is hard pressed to say that 'no collapse' has taken place and that we are still in a superposed state, since we have also gone from a bonding π electron orbital to an anti-bonding π* electron orbital state.

This going from one orbital to another, also means that we go from physics to chemistry or photochemistry, as demonstrated in the Grotthus-Draper first law of photochemistry, which states that light must be absorbed by a molecule in order for a photochemical reaction to take place [3]. The Stark-Einstein second law states that for each quantum of light absorbed by a chemical system, only one molecule is activated for a photochemical reaction [3]. The information carried by the photon has been passed on to the π* electron, which is now subject to the three main cascades of events.

If one still insists on clinging to a superposed state (which has now increased greatly in complexity), how do you deal with the next step in this process, which is the immediate conformational change of the retinal molecule from *cis* to *trans* in ~ 200 fs [4,5]? Do both molecular conformations remain in a superposed state also? In addition, there are a number of complex steps which take place immediately after this (as more fully covered in [2]) which would all have to be taken into consideration in the final superposed state, which supposedly then finally gets reduced *somewhere* in the visual cortex.

It is of interest to interject at this point, von Neumann's two-process projection postulate, and see if it might cast some light on the issue at hand [6]. The first process states that light or the measured quantum system $S$, interacts with a macroscopic measuring apparatus $M$ (the photoreceptor rod cell-rhodopsin molecule) for some physical quantity $Q$, with the interaction governed by the linear, deterministic Schrödinger equation [6,7]. The second process states that after this first stage of the measurement terminates, and one has a linear combination of products which are called entangled states, a second non-linear, indeterminate process takes place, the collapse of the wave packet.  It is important to mention here that process one and two are *temporally contiguous* [6,7].

In addition, the Bassi-Ghirardi theory faces another major problem, when one takes into consideration the fact that rhodopsin is the protein responsible for not only generating an optic nerve impulse in the visual receptors of our eyes but, also those found in all three phyla which possess eyes: mollusks, arthropods and vertebrates [8,9].  This amounts to over $1.3 \times 10^6$ *other living entities* which can collapse the wavefunction, the bulk of which are not *sentient* (I am guessing!).  And, whose mostly small sizes, in comparison with humans, would not allow the accumulation of a sufficient number of particles over such extremely short and varying distances and times, to permit reduction to take place under the QMSL theory.  I.e., it is nature itself that rules out the QMSL process in this animate setting!  The von Neumann projection postulate can more simply and directly accommodate *all* of these disparate entities without the need for so many intervening steps.

Finally, in order to clarify my position on this matter, I have been taking what is known as the internalist stance, as has been proposed by Matsuno [2, 10-11].  This means that the material act of distinguishing between before and after physical events, whatever they are, is taken to be most fundamental, irreducible and even ubiquitous inside this empirical world.  The linear approach, no matter how cherished it may be by the majority, would remain secondary at best.  Once one accepts this stance, nonlinearity intrinsic to the internal act of making distinctions, would turn out to be the rule rather than the exception, which will represent a new doctrine.